# Sociophysics: a personal testimony

Serge Galam

*Laboratoire des Milieux Désordonnés et Hétérogènes, Tour 13, Case 86, 4 place Jussieu, 75252 Paris Cedex 05, France*
CNRS UMR 7603

The origins of Sociophysics are discussed from a personal testimony. I trace back its history to the late seventies. My twenty years of activities and research to establish and promote the field are reviewed. In particular the conflicting nature of Sociophysics with the physics community is revealed from my own experience. Recent presentations of a supposed natural growth from Social Sciences are criticized.

## I. In the late seventies: the high rise of Statistical Physics

During the years 1975-80 Statistical Physics is blooming with the exact solving of the enigma of critical phenomena, one of the most resistant problems of physics. The so-called Modern theory of phase transitions with the renormalization group techniques and the epsilon expansion made condensed matter physics to enter its Golden Age [1].

The Mecca of this rebirth of Physics was at Cornell University in the US with all over the world several locations promptly established. Several hundreds of young physicists were entering the field with a great deal of excitement and a huge number of papers were published on the subject. Among the active spots was the physics department at Tel-Aviv University in Israel. And among the excited students was me.

After a doctorate at the University Pierre et Marie Curie in Paris I moved to Israel to complete a Ph.D. During my research I started to advocate the use of Modern theory of phase transitions to describe social, psychological, political and economical phenomena. My claim was motivated by an analysis of some epistemological contradiction within physics. On the one hand, the power of concepts and tools of Statistical Physics were enormous, and on the other hand, I was expecting that physics would soon reach the limits of investigating inert matter. On this basis to compensate the frustration, which would result from that contradiction, I was suggesting physicists to start applying Physics outside Physics. In particular to deal with the rich variety of behavior related to all kind of human activities. I published a long series of papers, several of them with P. Pfeuty, to legitimate and elaborate my suggested strategy of [2-9].

But such an approach was strongly rejected by almost every one, leading and non-leading physicists, young and old. To suggest humans could behave like atoms was look upon as a blaspheme to both hard science and human complexity, a total non-sense, something to be condemned. And it has been indeed condemned during fifteen years.

## II. The founding of Sociophysics

At that time I was aware of 2 papers along the spirit of my call for the creation of Sociophysics. One was by Callen and Shapiro in Physics Today [10] about fish bands, imitation and Ising spins. It was very enlightening but with no follow up. The other one was much more elaborated by Weidlich about the dynamics of opinion forming [11]. However it was more of an application of partial differential equations in the spirit of Volterra work than an application of the Modern theory of phase transitions.

On this basis, while keeping arguing with physicists about the possibility of Sociophysics, I wrote my first contribution to it in 1982 [12]. I applied Carnot principle of maximum entropy to societies to demonstrate that, contrary to the claimed ineluctability of the thermal death argument, there are no reasons to get pessimistic about it. However as for the above papers [10, 11] it was of a factual contribution and moreover written in French. Then, at the same time, with 2 other Ph.D. colleagues, Y. Gefen and Y. Shapir, we published a paper to state a global frame for Sociophysics as a new field of research [13]. In addition to modeling the process of strike in big companies using an Ising ferromagnetic model in an external reversing uniform field, the paper contains a call to the creation of Sociophysics. It is a manifesto about its goals, its limits and its danger. As such, it is the founding paper of Sociophysics [13], although it is not the first contribution per se to it.

It is worth to stress that we choose to submit our manuscript not to a physical journal, but to the Journal of Mathematical Sociology. It took over 2 years to get the paper accepted for publication after many exchanges with several referees. The paper appeared in 1982. But in the time being, we all have left Tel-Aviv University to pursue our physicist career as post doctorates in the US and I did not hear much about eventual reactions.

**III. The story behind the scene**

To illustrate the frame of mind of the physicists at that time with respect to the creation of Sociophysics it is very suggesting to mention what happened at the Department of Physics and Astronomy from Tel-Aviv University during the typing of our manifesto of Sociophysics [13]. First it must be recalled that the whole story occurred more than twenty years ago. At that time there were no personal computer and writing a paper was a real challenge with respect to have a typed manuscript with equations. You had to put your hand written work in line to the secretary who had the skill to do the job and no question of coming back and forth to make changes. And it is what we did, but once our paper was typed, ready to be sent to the Journal, the Chairman without notice, took it over and put it in a locked place.

Our manuscript has thus been sequestrated under the authority of the chairman and with the support of most of the faculty. It was a big scandal. We were denounced as putting at stake the department reputation of excellence while we were claiming our right to academic freedom. Then, the chairman denied academic freedom to us with the argument that we were not on tenured positions. In this context with our manuscript under arrest, A. Voronel, a former member of the refusenik seminar from Soviet Union and a tenured professor at Tel-Aviv University, endorsed our manuscript under his responsibility. Using his right to academic freedom, he allowed us to submit our paper for publication and we did [13].

Besides, its own interest, above story illustrates how hard was the opposition to Sociophysics from inside Physics. I could tell many more stories along this one which occurred during these fifteen years of personal and lonely fight to create and develop Sociophysics. Always the opposition arose due to the general frame in which I was positioning my various contributions. It was the idea of creating a new field of research within Physics to deal with human behaviors, which was disturbing the physicists, not a factual contribution which could always be looked upon as a marginal and exotic isolated event.

**IV. In place of playing tennis**

Although I was totally convinced of both the validity and the necessity of Sociophyics, I was not fool enough to jeopardize my academic career by putting all my research energy in it. Accordingly, to survive as a physicist, I kept doing orthodox physics, which at that time was also really exciting. I was presenting my Sociophysics activity as a "hobby". In place of playing tennis, I was playing Sociophysics. And as a matter of fact I did survive as a physicist. But it took me a lot of effort before I was able to produce another work to establish further the feasibility of Sociophysics.

After an additional paper using entropy in 1984 [14], it is only in 1986 that I was able to produce another significant contribution based on renormalization group concepts. I studied dictatorship effects induced by the use of the democratic rule of majority voting in hierarchical bottom-up elections [15]. I submitted my paper to the Journal of Mathematical Psychology and again it took two years of ongoing arguments with several referees before it was accepted. And after publication, as before I had again no feedback.

Few years latter while visiting the physics department at Tel-Aviv University, I met D. Stauffer who was yet doing only physics, but showed a real interest in my Sociophysics work. He encouraged me to submit a paper in the Journal of Statistical Physics, which I did. The paper was accepted in 1990 [16], yet with a letter from J. Lebowitz, the chief Editor, stressing he was accepting the paper because the referee reports were positive but that he personally did not believe at all in the validity of such an approach. And once again I got no feedback after publication. I later on published 3 more papers to extend my voting model [17, 19] with again no much reaction.

**V. The collaborating with a social scientist**

After all these efforts and contributions with no visible reaction, neither physicists not social scientists, I thought to reorient my strategy in seeking to collaborate directly with a social scientist along my using Physics to describe social behavior. While I was in New York, I met by chance S. Moscovici, a leading French social psychologist, who appeared rather interested in the adventure. We then start a very fruitful few years cooperation, which yielded a series of papers published during 1991-1995, most of them in the European Journal of Social Psychology [20-24]. But again, I got no feedback, except one invited paper in a book edited by a social scientist [25].

On this basis, I changed once more my strategy to come back to my initial postulate about the emerging of Sociophysics. It should come from physicists, so I started to publish my Sociophysics papers in physical journals (Thanks to Stauffer) [26]. And yet not much of a reaction came out of it. Nevertheless in the mid nineties some physicists were also turning "exotic", but to do Econophysics, with emphasize on analyzing financial data. Though this new trend was validating my earlier eighties prediction [2-9], I stayed isolated in my working field keeping on studying political and social behaviors [27].

**V. And the sun finally rose and failed**

And finally, more than fifteen after my first publications calling for Sociophysics [2-9], few additional physicists at last started to join along it. I was very happy to realize I was not crazy, or at least not the only one. However, most of these works either did not cite my earlier founding papers or did in a general manner along with other more recent works. Such a phenomenon being rather common in science, I sent some occasional mails mentioning my contributions to the relevant authors. But more recently few papers including Physical Review Letters presented "original" works, which were indeed reproducing exactly parts of my models. This prompted me to send more energetic mails, but I felt rather unease sounding like acting paranoiac so I cool off and stopped looking to cond-mat archives in too much details. However two very recent papers discussing the nature and origins of Sociophysics [28, 29], made me to react in writing with this very paper. It is not motivated only to release my "hurt ego", but mainly to restore the historical truth on the origins of Sociophysics since I am convinced it is a necessary condition to have Sociophysics to strength and established as a solid field of research. Especially to preserve the conflicting nature of Sociophysics which is an essential ingredient to it.

Accordingly it is of particular interest to note that both papers [28, 29] while different in the approach and style based the presentation of Sociophysics along a very politically correct view. They both create the illusion Sociophysics is a natural outgrowth of sociology, ignoring deliberately all of its very conflicting nature, with both the concerned scientists and the epistemological content. To support they idyllic view, they trace back its foundation to the work of Schelling [30] who according to them, was already doing Physics even without being aware of it. A significant "tour de passe-passe" to wrongly legitimate Sociophysics as a natural extension from social sciences as proven wrong in my testimony. One of them [28] relies strongly on Axelrod and Bennet [31] who developed a model for coalitions from Physics without mentioning I showed the approach was misleading [32], that is exactly the danger to be avoided with the kind of Sokal-Bricmont syndrome [33] when social scientists used Physics to justify their a priori theories.

**VI. Conclusion**

The paradoxical aspect of my testimony is the fact that physicists doing Sociophysics (me included) would like to have social scientists to get interested in their work. They complain social scientists do not read their contributions published in Physical Journals while they are all claiming an organic link to the 2 papers from social scientists [30, 31], papers most of them did not read. And at the same time, they do not read my earlier papers because either being published in non Physical Journals, they are not immediately at hands, or and they were published too many years ago while their memory is of only the last few months. Moreover while hoping for a real link from within social sciences, they mostly ignored my extended work in collaboration with a social scientist, S. Moscovici, a rather rare case.

All these facts should be interesting for the study of history of sciences. But at present, they shows we, as a community including myself, are much more interested in producing papers than really establishing a new field of research which could eventually turn helpful in solving some of the huge and major problems our societies are facing. But let us be optimistic and consider all these weaknesses are the direct outcome form the fact Sociophysics is still at its childhood.

**Poscriptum**

At this stage and in conclusion I feel to claim paternity over Sociophysics, even If I don't really know what such a claim would imply. Of course I am fully aware that to be the "father" does not mean to be the first one, as know in many aspects of life, and as clearly showed from the references I am giving. In addition, paternity always contains a little bit of a doubt, how to perform an ADN check. And on top of all, the question of the mother is open. So I am expressing here a feeling from my more than twenty years of fight for Sociophysics.

**References**

1) K. G. Wilson and J. Kogut, The renormalization group and the epsilon expansion, Phys. Reports 12C, 75-200 (1974)

2) S. Galam, Physicists as a revolutionary catalyst, Fundamenta Scientiae 1, 351-353 (1980)

3) P. Pfeuty and S. Galam, Les physiciens et la frustrations des électrons, La Recherche, (July-August 1981)

4) S. Galam, Sauver la nouvelle Byzance, La Recherche, Lettre, 127, 1320 (1981)

5) S. Galam and P. Pfeuty, Physicists are frustrated, Physics Today, Letter (April 1982)

6) S. Galam, Misère des physiciens, Pandore, 18, 57-58 (April 1982)

7) S. Galam, About imperialism of physics, Fundamenta Scientiae 3, 125 (1982)

8) S. Galam and P. Pfeuty, Should God save the queen? Physics Today, Letter (October 1983)

9) S. Galam and P. Pfeuty, Chaotic computer? Physics Today, Letter (October 1983)

10) E. Callen and D. Shapiro, A theory of social imitation, Physics Today, 23-28 (July 1974)

11) W. Weidlich, The statistical description of polarization phenomena in society, Br. J. math. statist. Psychol. 24, 251-266 (1971)

12) S. Galam, Entropie, désordre et liberté individuelle, Fundamenta Scientiae 3, 209-213 (1982)

13) S. Galam, Y. Gefen and Y. Shapir, Sociophysics: A mean behavior model for the process of strike, Journal of Mathematical Sociology 9, 1-13 (1982)

14) S. Galam, Entropy, Semiotext(e) 4, 73-74 (1984)

15) S. Galam, Majority rule, hierarchical structures and democratic totalitarianism: a statistical approach, Journal of Mathematical Psychology 30, 426-434 (1986)

16) S. Galam, Social paradoxes of majority rule voting and renormalization group, Journal of Statistical Physics 61, 943-951 (1990)

17) S. Galam, Political paradoxes of majority rule voting and hierarchical systems, International Journal of General Systems 18, 191-200 (1991)

18) S. Galam, Real space renormalization group and social paradoxes in hierarchical organisations, in Models of self-organization in complex systems (Moses), Akademie-Verlag, Berlin V.64, 53-59 (1991)

19) S. Galam, Paradoxes de la règle majoritaire dans les systèmes hiérarchiques, Revue de Bibliologie 38, 62-68 (1993)

20) S. Galam and S. Moscovici, Towards a theory of collective phenomena: Consensus and attitude changes in groups, European Journal of Social Psychology 21, 49-74 (1991)

21) S. Galam and S. Moscovici, Compromise versus polarization in group decision making, in Defense Decision Making, Springer-Verlag, 40-51 (1991)

22) S. Galam and S. Moscovici, A theory of collective decision making in hierarchical and non-hierarchical groups, Russian Psychological Journal 13, 93-103 (1993)

23) S. Galam and S. Moscovici, Towards a theory of collective phenomena: II. Conformity and power, European Journal of Social Psychology 24, 481-495 (1994)

24) S. Galam and S. Moscovici, Towards a theory of collective phenomena: III. Conflicts and forms of power, European Journal of Social Psychology 25, 217-229 (1995)

25) S. Galam, When humans interact like atoms, in Understanding group behavior, Vol. I, Chap. 12, 293-312, Davis and Witte, Eds., Lawrence Erlbaum Ass., New Jersey (1996)

26) S. Galam, Fragmentation versus stability in bimodal coalitions, Physica A230, 174-188 (1996)

27) S. Galam, Rational group decision making: a random field Ising model at T=0, Physica A238, 66-80 (1997)

28) P. Ball, Utopia theory, Physics World (October 2003)

29) D. Stauffer, Introduction to Statistical Physics outside Physics, Physica A, this issue

30) T.C. Schelling, J. Mathematical Sociology 1, 143 (1971)

31) R. Axelrod and D. S. Bennett, A landscape theory of aggregation, Br. J. Political Sci. 23, 211-233 (1993)

32) S. Galam, Comment on A landscape theory of aggregation, Br. J. Political Sci. 28, 411-412 (1998)

33) A. Sokal and J. Bricmont , Intellectual Impostures, Profile Books Ltd, London (1999)